\documentstyle{article}
\makeatletter
\def\faketitle{
\begin{center}
   {\Large\sc \@title \par}        % Set title in \Large size.
   \vskip 2em                  % Vertical space after title.
   {\large\sc   \@firstauthors \par}
   \vskip 1em                  % Vertical space after author.
   {\it \@firstaddress\\[.5em] }
   \vskip 1em                  % Vertical space after title.
   {\large\sc   \@secondauthors \par}
   \vskip 1em                  % Vertical space after author.
   {\it \@secondaddress\\[.5em] }
   \vskip 1em                  % Vertical space after title.
\end{center} \par
}

\def\maketitle{
\begin{center}
   {\Large\sc \@title \par}        % Set title in \Large size.
   \vskip 2em                  % Vertical space after title.
   {\large\sc   \@firstauthors \par}
   \vskip 1em                  % Vertical space after author.
   {\it \@firstaddress\\[.5em] }
   \vskip 1em                  % Vertical space after title.
\end{center} \par
}

\def\firstaddress#1{\def\@firstaddress{#1}}
\def\firstauthors#1{\def\@firstauthors{#1}}

\twocolumn
\advance\textheight by \topskip
\def\section{\@startsection {section}{1}{\z@}{-3.5ex plus -1ex minus
    -.2ex}{2.3ex plus .2ex}{\bf }}
\def\subsection{\@startsection{subsection}{2}{\z@}{-3.25ex plus -1ex minus
   -.2ex}{1.5ex plus .2ex}{\it }}

\def\@makefnmark{\hbox to 0pt{$^{\@thefnmark}$\hss}}

\renewenvironment{thebibliography}[1]
	{\begin{list}{\arabic{enumi}.}
	{\usecounter{enumi}\setlength{\parsep}{0pt}
	 \setlength{\itemsep}{0pt} \settowidth
	{\labelwidth}{#1.}\sloppy}}{\end{list}}

\newcounter{arabiclistc}

\def\@citex[#1]#2{\if@filesw\immediate\write\@auxout
	{\string\citation{#2}}\fi
\def\@citea{}\@cite{\@for\@citeb:=#2\do
	{\@citea\def\@citea{,}\@ifundefined
	{b@\@citeb}{{\bf ?}\@warning
	{Citation `\@citeb' on page \thepage \space undefined}}
	{\csname b@\@citeb\endcsname}}}{#1}}

\newif\if@cghi
\def\cite{\@cghitrue\@ifnextchar [{\@tempswatrue
	\@citex}{\@tempswafalse\@citex[]}}
\def\citelow{\@cghifalse\@ifnextchar [{\@tempswatrue
	\@citex}{\@tempswafalse\@citex[]}}
\def\@cite#1#2{{$\null^{#1}$\if@tempswa\typeout
	{IJCGA warning: optional citation argument
	ignored: `#2'} \fi}}

\def\baselinestretch{1.0}
\ifx\selectfont\undefined
\@normalsize\else\let\glb@currsize=\relax\selectfont
\fi

\ifx\selectfont\undefined
\def\@singlespacing{%
\def\baselinestretch{1}\ifx\@currsize\normalsize\@normalsize\else\@currsize\fi%
}
\else
\def\@singlespacing{\def\baselinestretch{1}\let\glb@currsize=\relax\selectfont}
\fi

\long\def\@makecaption#1#2{
   \vskip 10pt
   \setbox\@tempboxa\hbox{\footnotesize #1: #2}
   \ifdim \wd\@tempboxa >\hsize   % IF longer than one line:
       \leftskip 0pt plus 1fil
       \rightskip 0pt plus -1fil
       \parfillskip 0pt plus 2fil
       \footnotesize #1: #2\par   %   THEN set as ordinary paragraph.
     \else                        %   ELSE  center.
       \hbox to\hsize{\hfil\box\@tempboxa\hfil}
   \fi}
\makeatother
\begin{document}

\bibliographystyle{unsrt}    % for BibTeX - sorted numerical labels

\newcommand{\lm}[1]{\mbox{$\lambda_{#1}$}}
\newcommand{\m}[1]{\mbox{$m_{#1}^{2}$}}
\newcommand{\p}[1]{\mbox{$\phi_{#1}$}}
\renewcommand{\H}[1]{\mbox{$H_{#1}$}}
\newcommand{\Hd}[1]{\mbox{$H_{#1}^{\dag}$}}
\newcommand{\s}{\mbox{supersymmetry }}
\renewcommand{\ss}{\mbox{supersymmetric }}
\newcommand{\e}{\mbox{electroweak }}

\def\ap#1#2#3   {{\em Ann. Phys. (NY)} {\bf#1} (#2) #3.}
\def\apj#1#2#3  {{\em Astrophys. J.} {\bf#1} (#2) #3.}
\def\apjl#1#2#3 {{\em Astrophys. J. Lett.} {\bf#1} (#2) #3.}
\def\app#1#2#3  {{\em Acta. Phys. Pol.} {\bf#1} (#2) #3.}
\def\ar#1#2#3   {{\em Ann. Rev. Nucl. Part. Sci.} {\bf#1} (#2) #3.}
\def\cpc#1#2#3  {{\em Computer Phys. Comm.} {\bf#1} (#2) #3.}
\def\err#1#2#3  {{\it Erratum} {\bf#1} (#2) #3.}
\def\ib#1#2#3   {{\it ibid.} {\bf#1} (#2) #3.}
\def\jmp#1#2#3  {{\em J. Math. Phys.} {\bf#1} (#2) #3.}
\def\ijmp#1#2#3 {{\em Int. J. Mod. Phys.} {\bf#1} (#2) #3.}
\def\jetp#1#2#3 {{\em JETP Lett.} {\bf#1} (#2) #3.}
\def\jpg#1#2#3  {{\em J. Phys. G.} {\bf#1} (#2) #3.}
\def\mpl#1#2#3  {{\em Mod. Phys. Lett.} {\bf#1} (#2) #3.}
\def\nat#1#2#3  {{\em Nature (London)} {\bf#1} (#2) #3.}
\def\nc#1#2#3   {{\em Nuovo Cim.} {\bf#1} (#2) #3.}
\def\nim#1#2#3  {{\em Nucl. Instr. Meth.} {\bf#1} (#2) #3.}
\def\np#1#2#3   {{\em Nucl. Phys.} {\bf#1} (#2) #3.}
\def\pcps#1#2#3 {{\em Proc. Cam. Phil. Soc.} {\bf#1} (#2) #3.}
\def\pl#1#2#3   {{\em Phys. Lett.} {\bf#1} (#2) #3.}
\def\prep#1#2#3 {{\em Phys. Rep.} {\bf#1} (#2) #3.}
\def\prev#1#2#3 {{\em Phys. Rev.} {\bf#1} (#2) #3.}
\def\prl#1#2#3  {{\em Phys. Rev. Lett.} {\bf#1} (#2) #3.}
\def\prs#1#2#3  {{\em Proc. Roy. Soc.} {\bf#1} (#2) #3.}
\def\ptp#1#2#3  {{\em Prog. Th. Phys.} {\bf#1} (#2) #3.}
\def\ps#1#2#3   {{\em Physica Scripta} {\bf#1} (#2) #3.}
\def\rmp#1#2#3  {{\em Rev. Mod. Phys.} {\bf#1} (#2) #3.}
\def\rpp#1#2#3  {{\em Rep. Prog. Phys.} {\bf#1} (#2) #3.}
\def\sjnp#1#2#3 {{\em Sov. J. Nucl. Phys.} {\bf#1} (#2) #3.}
\def\spj#1#2#3  {{\em Sov. Phys. JEPT} {\bf#1} (#2) #3.}
\def\spu#1#2#3  {{\em Sov. Phys.-Usp.} {\bf#1} (#2) #3.}
\def\zp#1#2#3   {{\em Zeit. Phys.} {\bf#1} (#2) #3.}

\setcounter{secnumdepth}{2} % Number sections and subsections

%%%%%%%%%%%%%%%%%%%%%%%%%%%%%%%%%%%%%%%%%%%%%%%%%%
%                                                %
%    BEGINNING OF TEXT                           %
%                                                %
%%%%%%%%%%%%%%%%%%%%%%%%%%%%%%%%%%%%%%%%%%%%%%%%%%

\title{ELECTROWEAK BARYOGENESIS IN A SUPERSYMMETRIC MODEL\\
To appear in HEP95 Proceedings, Brussels, 1995.}

\firstauthors{A. T. Davies, C. D. Froggatt, G. Jenkins, R. G. Moorhouse }

\firstaddress{Department of Physics and Astronomy,University of Glasgow,
Glasgow G12 8QQ, U.K.}

%\secondauthors{ A.N. Other }

%if there are no second authors then comment out the line and adjust the
%maketitle command and \def\secondauthor... command in snow.sty

%\secondaddress{Department of Exotic Results, University of Heavens,
%Universe Road 999,\\  Whoknowswhere ZZZ123, Paradise}

%if there are no second authors then comment out the line and adjust the
%maketitle command and \def\secondaddress... command in snow.sty
\twocolumn[\maketitle]

\section{Introduction and Formalism of the NMSSM}
It is well known that there is
difficulty in
sustaining the hypothesis \cite{Cohn,Dine} of baryogenesis at the electroweak
 phase transition in the minimal standard model.  To overcome these
difficulties
attention has been given to extensions of the minimal standard model, involving
the addition of extra scalars \cite{Andn,Cohn,Davl}.
Prominent among these is the minimal supersymmetric standard model, MSSM, where
the Higgs sector is just two doublets \cite{Gun2,Espa}. Here we shall discuss,
 with a perturbative treatment, the next-to-minimal model, NMSSM, which has
additionally one singlet Higgs scalar \cite{Gun2}.In the absence of hard
information we have to adopt a hypothesis on the SUSY
breaking scale and on the spectrum of the particles, and ours is the simplest
possible. We follow a number of papers of recent years in taking the SUSY
breaking scale,$M_S$, to be of the order of 1 TeV; we take perfect
supersymmetry above that scale. Then at $M_S$ the quartic scalar couplings are
fixed by the gauge couplings and  two more parameters. We then use the
renormalization group equations to run down the quartic couplings to
the electroweak scale, where we investigate the nature of the phase change.
There are also cubic and quadratic supersymmetry breaking couplings,and there
results a space of variable parameters in which we investigate what proportion
leads to a first order electroweak phase change, and so is compatible with
electroweak baryogenesis.

There has been quite considerable previous work on the electroweak phase change
in the MSSM. We are not aware of so much on the
NMSSM . The work of Pietroni\cite{Piet} has pointed up that the NMSSM, in
contrast to the MSSM, has cubic terms in the scalar field potential at tree
level leading to the possibility of a potential barrier in radial directions
even at tree level. That work uses a unitary gauge which we consider to be
not so secure a basis for the consideration of phase changes as the Landau
gauge which we use \cite{Doln,Davl}.

We start with the tree level potential,which is \cite{Gun2}
\begin{eqnarray*}
 \lefteqn{V_{0}  = }
 & & \frac{1}{2}(\lm{1}(\Hd{1}\H{1})^2 +\lm{2}(\Hd{2}\H{2})^2) + \\
 & & (\lm{3}+\lm{4})(\Hd{1}\H{1})(\Hd{2}\H{2}) -
\lm{4}\left| \Hd{1}\H{2}\right|^{2}+ \\
 & & (\lm{5}\Hd{1}\H{1} +\lm{6}\Hd{2}\H{2})N^{\star}N+
(\lm{7}\H{1}\H{2}N^{\star2}+hc) + \\
 & & \lm{8}(N^{\star}N)^2+
(\left|\mu\right|^2+(\lambda\mu^{\star}N+hc))(\Hd{1}\H{1}+\Hd{2}\H{2}) \\
 & & + \m{1}\Hd{1}\H{1}+\m{2}\Hd{2}\H{2} + \m{3}N^{\star}N-\\
 & & ((m_4\H{1}\H{2}N+
\frac{1}{3}m_5N^3-\frac{1}{2}\m{6}\H{1}\H{2}-\m{7}N^2)+hc)
\end{eqnarray*}
where $\H{1}^{T} = (\H{1}^{0},\H{1}^{-})$, $\H{2}^{T} = (\H{2}^{+},\H{2}^{0})$,
and $\H{1}\H{2} = \H{1}^{0}\H{2}^{0}- \H{1}^{-}\H{2}^{+}$.
The terms involving $\mu$ arise from the $\mu$ term in the superpotential. The
last two lines comprise all possible soft supersymmetry breaking terms
\cite{Gun2}.
$V_{0}$ is a function of 10 real scalar fields, \mbox{$\phi =
\p{1},\p{2},\ldots,\p{10}$}, 4 for each of the Higgs doublets and 2 for the
singlet, N.

For
simplicity, and to automatically ensure real VEVs, we shall follow the usual
practice and take the parameters real.
The boundary values at $M_S$ of the quartic couplings are given by \cite{Ekw1}
$$\lm{1} = \lm{2} = \frac{1}{4}(g_2^2+g_1^2),
 \lm{3} = \frac{1}{4}(g_2^2-g_1^2), \lm{4} = \lambda^2-\frac{1}{2}g_2^2 ,$$

$$\lm{5}=\lm{6}=\lambda^2,\lm{7}=-{\lambda}k,\lm{8}=k^2 $$
and are developed down to $M_{Weak}$ by using the appropriate RG equations
\cite{Ekw1}. $\lambda$ and k, from the superpotential, are free parameters at
$M_S$\cite{Gun2}. However they are linked to the one important Yukawa coupling
 \footnote{we are not considering large $tan\beta$ here}
, $g_t$, by 3 simultaneous RG equations \cite{Ekw1}; in developing from high
energy down to $M_S$ their values there should not be such that they
correspond to divergent or unnaturally large values at high energy. The
$\m{1},\m{2},\m{3}$ are standard mass parameters and are to be specified in
terms of the VEVs and other parameters by the usual requirement that
 $V_{0}(\phi)$, \mbox{$\phi =
\p{1},\p{2},\ldots,\p{10}$}, with the parameters \lm{i} assumed renormalised
at the \e scale, be a minimum at the neutral VEVs:
\begin{equation} \label{eq.vevs} \langle\H{1}\rangle
= \left( \begin{array}{c} v_{1}\\ 0 \end{array}
\right), \; \langle\H{2}\rangle = \left( \begin{array}{c} 0\\
v_{2} \end{array} \right),\langle N\rangle=x
\end{equation}
where $v_{1}$, $v_{2}$ and x are real and
\mbox{$v = \sqrt{v_{1}^{2} + v_{2}^{2}} = 174GeV$}. The scalar mass-squared
matrix gives rise to 7 massive physical particles and 3 zero mass would-be
 Goldstone bosons.We can now discuss the other parameters in $V_0$.

Firstly there are the terms involving $\mu$ which arise from the $\mu$ term in
 the superpotential. This raises the mu-problem(first noted in the MSSM)
\cite{Absw}; $\mu$ would naturally be expected to take on a value of the order
of
magnitude of the fundamental scale of the theory, whereas phenomenologically it
should be of
the order of the other \e terms. We do not take the point of view that the
NMSSM can solve this by its having largely phenomenologically equivalent terms
in the N field and simply setting $\mu=0$ \cite{Ekw1}. This can
have its own difficulties when a resulting $Z_3$ symmetry gives rise to domain
walls\cite{Absw}. We tolerate the mu-problem. It should be noted that we have
extra $Z_3$ breaking by the phenomenological term $\m{6}\H{1}\H{2}$. Secondly
 there are the remaining soft parameters $m_4,m_5,\m{7}$; their provenance
as completing the most general NMSSM breaking $V_0$ was given in
 Ref.\cite{Gun2}.

Going now to the T-dependent terms in the effective potential we proceed in the
usual way by the loop expansion. For a temperature sufficiently high compared
 to the masses we use the well known expansion in descending powers of
 temperature down to the logarithmic term.
We include the 1-loop contribution of the Higgsinos in
the case where the masses of the winos and binos are taken large, of order
$M_S$. The size of the parameters gives
these Higgsinos masses to be significantly less than $M_S$ and thus they
 should be taken into account; the only quark included is the top. We also
modify the 1-loop calculation by including ring diagrams, and suppressing
gauge boson longitudinal polarization excitations \cite{Dine}. For smaller T
we use a Boltzmann suppressed form\cite{Andn} for the effective potential.

\section{Results}
Calculations with the kinetic equations for the dilution of
baryonic charge just after the phase transition give a baryon preservation
condition \cite{Cohn,Dine}
\begin{equation} \label{eq.voverT}
\frac{v(T_{crit})}{T_{crit}} \geq \xi
\end{equation}
where $\xi$ varies between about 0.9 and 1.5 according to the gauge and
other couplings of the theory.
We take $\xi = 1$. In practice in various theories just two criteria have
been used to find the critical temperature, $T_{crit}$, and authors have
 made a choice of either one or the other; we use both and compare. One is the
 temperature, $T_0$, at which, for decreasing T, the curvature of the
effective potential $V(\phi,T)$ at $\phi=0$(assumed to be the previous global
minimum) first vanishes in the Higgs doublet neutral field directions. The
other is the temperature,$T_C$, at which the value of V at a minimum with a
markedly non-zero $v'$ first becomes the global minimum of $V( v_{1}',v_{2}',
x',T)$. We note that the term in $\mu$ in $V_0$ gives rise to a term in
$V\propto T^{2} \lambda \mu N$ whose existence means that the origin cannot be
a minimum. Thus only for $\mu=0$ is the first criterion,$T_{crit}=T_0$,
applicable. Now the shape of V depends on the theory parameters:-
,$[\mu, \lambda, k]$; $[v=174GeV,tan\beta=\frac{v_2}{v_1},x]$ (replacing
\m{1},\m{2},\m{3}); $[M_{ch}]$ (mass of charged Higgs, replacing $m_4$);
 $[m_5, \m{6}, \m{7}]$.

In the work reported here we have adopted the values of Ref.(\cite{Ekw1})
in taking $k=.1, \lambda=.65$. For other parameters we have searched in the
 regions $1 < tan\beta < 3, 200 < M_{ch} < 300GeV, -5 < \frac{\m{6}}{50^2} < 5
, -5 < \frac{\m{7}}{50^2} < 5, -1 < \frac{m_5}{50} < 1 $, while for x we
have used x=174GeV, having found that values significantly bigger or smaller
 greatly restricted the range of the other parameters compatible with the T=0
criteria. Considerations on the special parameter $\mu$ are given below,
where we now outline our current results:

1. $\mu=0$: A search over a grid of 200,000 sets of parameters found a basis
space of about 20,000 giving an acceptable broken T=0 electroweak vacuum.
The curvature criterion gave about 10\% of this basis space compatible with
 baryon number preservation with $T_{crit}=T_{0}$ mostly in the range
50-150 GeV. The lightest Higgs scalar was of the order of 100 GeV, and the
lightest Higgsino was of a similar mass.

Around 80\% of these acceptable cases were also acceptable on the equal
minimum criterion,$T_{crit}=T_{C}$, though with a somewhat different
$T_{crit}$, the difference being as much as 50 GeV in a few cases. On the
average
$T_{C}$ is 10 GeV greater than $T_{0}$.

2. $\mu \ne 0$: In the region close to the previous successful cases, sets of
 parameters exist with values of $\mu$ of magnitude up to 40 GeV
 with negative values preferred.

3. Conclusions: We confirm, and quantify, previous results that electroweak
baryon preservation is compatible with the NMSSM and we additionally
find that a $\mu$ parameter of moderate magnitude is acceptable.

The investigation of the parameter space is not complete and is continuing.

\end{document}